%% file: cnovit.tex
\documentclass[oldversion]{aa}
\usepackage{amsmath,graphics,amssymb}

\usepackage{natbib}
\defcitealias{nltei}{Paper I}
\defcitealias{bezvi}{Paper II}

\newcommand{\mdot}{\ensuremath{\dot M}}
\newcommand{\Teff}{\ensuremath{T_\mathrm{eff}}}
\newcommand{\zav}[1]{\left(#1\right)}
\newcommand{\ms}{\ensuremath{\text{M}_{\odot}}}

\newcommand{\msr}{\ensuremath{\ms\,\text{year}^{-1}}} 
\newcommand{\vel}{{v}}
\newcommand{\vinfty}{\ensuremath{\vel_\infty}}
\newcommand{\hzav}[1]{\left[#1\right]}
\newcommand{\rs}{\ensuremath{R_*}}
\newcommand\de{\text{d}}
\newcommand\tlustocerv{\mathrm{thick}\atop\mathrm{lines}}

\allowdisplaybreaks

\begin{document}

\title{CNO-driven winds of hot first stars}

\author{J.  Krti\v{c}ka\inst{1} \and J. Kub\'at\inst{2}}
\authorrunning{J. Krti\v{c}ka and J. Kub\'at}

\institute{\'Ustav teoretick\'e fyziky a astrofyziky P\v{r}F MU,
            CZ-611 37 Brno, Czech Republic, \email{krticka@physics.muni.cz}
           \and
           Astronomick\'y \'ustav, Akademie v\v{e}d \v{C}esk\'e
           republiky, CZ-251 65 Ond\v{r}ejov, Czech Republic}

\date{Received:}

\abstract{During the evolution of first stars, the CNO elements may emerge on
their surfaces due to the mixing processes. Consequently, these stars may have
winds driven purely by CNO elements. We study the properties of such stellar
winds and discuss their influence on the surrounding environment. 
%{\bf For this
%purpose we used our own NLTE models and tested for which stellar parameters
%Ed tomu obdelnicku nerozumim
%Kued: ja taky ne, ale to, cos napsal, dava smysl
%that correspond to the first stars in different evolutionary stages the CNO winds
%may exist.}
For this purpose, we used our own NLTE models and tested which stellar
parameters of the first stars at different evolutionary stages result in CNO
winds.
If such winds are possible, we calculate their hydrodynamic structure
and predict their parameters. We show that, while the studied stars do not have
any wind driven purely by hydrogen and helium, CNO driven winds exist in more
luminous stars. On the other hand, for very hot stars, CNO elements are too
ionized to drive a wind. In most cases the derived mass-loss rate is much
less than calculated with solar mixture of elements.
%{\bf This is because
%Ed: tady jsem to upravil jinak
%Kued: Me se ten jeji navrh  docela libi, ale takhle je to snad taky ok.
%elements heavier than CNO influence the wind mass-loss rate most in the present
%hot stars.}
%most important ones that influence the wind mass-loss rate.
This is because wind mass-loss rate in present hot stars is dominated by
elements heavier than CNO.
We conclude that,
until a sufficient amount of these elements is created, the influence of
line-driven winds is relatively small on the evolution of hot stars (which are not close to the
Eddington limit).

}

\maketitle

\section{Introduction}

The only elements created in non negligible amounts during primordial
nucleosynthesis are helium and hydrogen \citep [e.g.,][] {coc}. Consequently,
first stars, which formed from the pristine gas processed during this primordial
nucleosynthesis, were pure hydrogen-helium stars. Numerical simulations of first
star formation show that the primordial nucleosynthesis not only determined the
chemical composition of these stars, but the absence of heavier elements
also influenced the initial mass function of these stars
\citep[e.g.,][]{bcl,naum}. In the absence of heavier elements, the only efficient
cooling processes during the collapse of primordial haloes are those of
molecular hydrogen. Since the molecular hydrogen line cooling is less efficient
than, say cooling from dust, the temperature in the first star-forming
regions was much higher than in the present-day star-forming regions. In the
%Ed tady jsem pridal to a smazal typical
absence of fragmentation this led to very high masses of first
stars (of the order of $10-100\,\ms$, \citealt{ompal}).

The general picture of the evolution of first stars is in many aspects different
from the evolution of present hot stars \citep{bezmari,klap,samhir}. One of the
most important differences is the possible existence of pair-instability supernovae,
which create a very typical pattern of nuclear yields \citep{parne}. The
possibility of creating a pair-instability supernova is closely connected to the
total amount of mass lost by an individual star during its evolution. If the
star does lose a significant fraction of the mass during its evolution, it may avoid
the pair-instability supernova eruption \citep{sylvie}.

Current massive stars lose their mass via line-driven winds \citep[see] [for
reviews] {kupul,kkpreh}. These winds are accelerated mainly due to the
absorption in the lines of heavier elements, such as iron, carbon, nitrogen, and
oxygen. For these stars the amount of mass lost by the star per unit of time
(the mass-loss rate) can be in principle derived both from observations and from
theoretical modelling \citep[e.g.,][]{pulamko}. However, given that there have
not been any available observations of first stars up to now, we have to rely on
the theoretical predictions alone in the case of first stars.

Stellar winds of pure hydrogen-helium extremely massive first stars were studied
by \citet[hereafter \citetalias{bezvi}]{bezvi}. We showed that the homogeneous
line-driven winds of these stars are unlikely and that only an extremely weak
pure hydrogen wind could be possible for stars very close to the Eddington
limit. Stellar winds of very low-metallicity massive stars were studied by
\citet{kudmet} assuming the solar mixture of elements. On the basis of these
models, \citet{smari} concludes that line-driven winds influence the evolution
of non-rotating, very low-metallicity stars only negligibly. Stellar winds of
stars at extremely low metallicity might be possible only if the star is very
close to the Eddington limit \citep{kudmet,vikowr,grahamz}.

During the core He-burning phase of first stars, primary nitrogen emerges on the
stellar surface \citep{mee,samhir}. This primary nitrogen is synthetised in the
H-burning shell due to the rotational mixing of carbon and oxygen produced in
the helium core. Moreover, evolutionary models of massive first stars
\citep[e.g.,][]{mee,samhir} show that subsequent generation of stars could
contain relatively large amounts of CNO elements, whereas the abundance of
iron-peak elements was very low. This picture is also supported by observations
of carbon-rich, extremely metal-poor stars \citep[e.g.,][]{pannorris}.

However, the detailed study of stellar winds driven purely by CNO elements is
still missing. Here we study such winds for the parameters typical of hot first
stars.

%%%%%%%%%%%%%%%%%%%%%%%%%%%%%%%%%%%%%%%%%%%%%%%%%%%%%%%%%%%%%%%%%%%%%%%%
\section{Description of models}

\subsection{Basic assumptions}

The models used in this paper are based on the NLTE wind models of \citet[hereafter
\citetalias{nltei}]{nltei}. Here we only summarise their basic features and
describe improvements in atomic data.

We assume spherically symmetric stationary stellar wind. The excitation and
ionization state of the considered elements is derived from the statistical
equilibrium (NLTE) equations. For some ions we improved the atomic data and
adopted the ionic models from the OSTAR2002 grid of model stellar atmospheres
\citep[see Table~\ref{prvky}] {ostar2003,bstar2006}. These ionic models are
based mainly on the Opacity Project data \citep{topp,top1,toptu,topt,topf}. For
some of these stars, the ionization fractions of some ions in the envelopes are so
low that it is not meaningful to include these ions in the solution of the NLTE
equations. The conditions for including such ions are also given in
Table~\ref{prvky}. The ionic list is modified in such a way that
the highest considered ion is only included as its ground state.

\begin{table}[t]
\caption{Atoms and {ions} included in the NLTE calculations.}
\label{prvky}
{\centering
\begin{tabular}{lccc}
\hline
Ion & Levels$^1$ & Data & Comment$^2$\\
\hline
\ion{H}{i}   &  9& \\
\ion{H}{ii}  &  1& \\
\ion{He}{i}  & 14& & $T_\text{eff}<70\,000\,\text{K}$\\
\ion{He}{ii} & 14& \\
\ion{He}{iii}&  1& \\
\ion{C}{i}   & 26& & $T_\text{eff}<30\,000\,\text{K}$ \\
\ion{C}{ii}  & 14& & $T_\text{eff}<35\,000\,\text{K}$ \\
\ion{C}{iii} & 23& OSTAR2002$^3$ & $T_\text{eff}<70\,000\,\text{K}$ \\
\ion{C}{iv}  & 25& OSTAR2002\\
\ion{C}{v}   & 11&\\
\ion{C}{vi}  & 10& & $T_\text{eff}>30\,000\,\text{K}$\\
\ion{C}{vii} &  1& & $T_\text{eff}>50\,000\,\text{K}$ \\
\ion{N}{i}   & 21& & $T_\text{eff}<30\,000\,\text{K}$\\
\ion{N}{ii}  & 14&& $T_\text{eff}<35\,000\,\text{K}$ \\
\ion{N}{iii} & 32& OSTAR2002 & $T_\text{eff}<70\,000\,\text{K}$\\
\ion{N}{iv}  & 23& OSTAR2002\\
\ion{N}{v}   & 16& OSTAR2002\\
\ion{N}{vi}  & 15& & $T_\text{eff}>20\,000\,\text{K}$\\
\ion{N}{vii} &  1& & $T_\text{eff}>30\,000\,\text{K}$\\
\ion{O}{i}   & 12& & $T_\text{eff}<30\,000\,\text{K}$\\
\ion{O}{ii}  & 50&& $T_\text{eff}<35\,000\,\text{K}$ \\
\ion{O}{iii} & 29& OSTAR2002 & $T_\text{eff}<70\,000\,\text{K}$\\
\ion{O}{iv}  & 39& OSTAR2002\\
\ion{O}{v}   & 14&\\
\ion{O}{vi}  & 20& OSTAR2002 & $T_\text{eff}>20\,000\,\text{K}$\\
\ion{O}{vii} &  1& & $T_\text{eff}>30\,000\,\text{K}$\\
\hline
\end{tabular}}

$^1$ An individual level or a set of levels merged into a superlevel.\\
$^2$ Note: the comment (if any) shows for which stars the
individual ion is taken into an account.
If the comment is missing, then the ion is considered for all models.\\
$^3$ The ionic model taken from \citet{ostar2003}.
\end{table}

As in \citetalias{nltei}, the radiative transfer problem is artificially split
into two parts, namely the radiative transfer in continuum and the radiative
transfer in lines. The solution of the radiative transfer equation in continuum
is based on the Feautrier method in the spherical coordinates
\citep{sphermod,dis} with inclusion of all free-free and bound-free transitions
of model ions, however neglecting line transitions.

The radiative transfer in lines is solved in the Sobolev approximation
\citep{cassob} neglecting continuum opacity and line overlaps. The radiative
force is calculated in the Sobolev approximation using data from the VALD
database \citep{vald1,vald2}. To test the completeness of our line list, we
compared these data with the line-lists provided by the NIST database
\citep{nist} and given by \citet{kur01} and enlarged the original data set when
necessary. The radiative force due to the light scattering on free electrons is
also included in the model calculations.

The surface emergent flux (i.e., the lower boundary condition for the radiative
transfer in wind) is taken from H-He spherically symmetric NLTE model stellar
atmospheres of \citet[and references therein]{kub}.

\begin{table}[t]
\caption{Radius $R_*$, mass $M$, and the effective temperature $T_\text{eff}$
of studied model stars}
\label{hvezpar}
%\centering
\begin{tabular}{ccccc@{}cccc}
\multicolumn{4}{c}{ZAMS stars} & &
\multicolumn{4}{c}{Evolved stars}\\
\cline{1-4}
\cline{6-9}\\[-3mm]
\cline{1-4}
\cline{6-9}
Model & $R_*$ & $M$ & $T_\text{eff}$ & & Model & $R_*$ & $M$ & $T_\text{eff}$ \\
& $[\text{R}_\odot]$ & $[\text{M}_\odot$] & [kK] & &
& $[\text{R}_\odot]$ & $[\text{M}_\odot$] & [kK] \\
\cline{1-4}
\cline{6-9}
M999& 4.23 & 100 & 94.4 & & M999-0 &  8.2 & 100 & 73.6 \\
M700& 3.44 &  70 & 89.5 & & M999-1 & 56.4 & 100 & 29.9 \\
M500& 2.82 &  50 & 84.1 & & M999-2 &125   & 100 & 20.1 \\
M300& 2.10 &  30 & 74.0 & & M999-3 &510   & 100 & 10.0 \\
M200& 1.65 &  20 & 65.3 & & M500-1 & 11.1 &  50 & 50.0 \\
M150& 1.48 &  15 & 57.3 & & M500-2 & 33.7 &  50 & 29.9 \\
M120& 1.42 &  12 & 49.9 & & M500-3 & 72.0 &  50 & 20.6 \\
M100& 1.37 &  10 & 44.5 & & M500-4 & 303  &  50 & 10.1 \\
M090& 1.34 &   9 & 41.6 & & M200-1 &  4.1 &  20 & 50.0 \\
M080& 1.31 &   8 & 38.5 & & M200-2 & 19.9 &  20 & 24.5 \\
M070& 1.30 &   7 & 34.8 & & M100-1 & 11.4 &  10 & 20.2 \\
M060& 1.27 &   6 & 31.4 & & M100-2 & 45.6 &  10 &  9.8 \\
M050& 1.23 &   5 & 27.7 & & M050-1 &  5.1 &   5 & 20.1 \\
M040& 1.17 &   4 & 23.6 \\ 
M030& 1.11 &   3 & 19.1 \\ 
M020& 1.00 &   2 & 13.7 \\ 
\cline{1-4}
\cline{6-9}
\end{tabular}
\end{table}

Parameters of these stars (given in Table~\ref{hvezpar}) were obtained
according to the evolutionary calculations of initially zero-metallicity stars
derived by \citet{bezmari}. We selected stellar parameters corresponding to the
zero-age main sequence of these models and to later evolutionary phases to cover
a larger area of the HR diagram.

%-----------------------------------------------------------------------
%Ed: upraveno
\subsection{The wind tests}

Since we do not know in advance whether the radiative force is strong enough for a given star and chemical composition to drive a wind,
we first test whether a wind may even exist. For these tests \citepalias[see
also][]{bezvi}, the hydrodynamical variables (velocity, temperature, and the
density) are kept fixed. This enables us to calculate the model occupation
numbers and the radiative force
even in the case when the wind does not exist.
The
comparison of calculated radiative force and the gravitational force serves as a
test of whether the wind exists; i.e., the winds are only possible when the magnitude of
the radiative acceleration $g^\text{rad}$ is greater than the magnitude of the
gravity acceleration $g$,
\begin{equation}
\label{trebova}
g^\text{rad}>g.
\end{equation}
Besides the cotribution of lines, we also include
the radiative acceleration due to the bound-free and free-free
transitions in the calculation of the total radiative acceleration
$g^\text{rad}$.

The velocity structure of our models is given by an artificial velocity law
\citepalias[see the discussion in][]{bezvi}
\begin{equation}
v(r)=10^{-3}\sqrt{\frac{5}{3}\frac{kT_{\text{eff}}}{m_\text{H}}}+
2 \times 10^{8}\,\text{cm\,s}^{-1}\frac{r-R_*}{R_*},
\end{equation}
where $T_\text{eff}$ is the stellar effective temperature, $R_*$ the stellar
radius, and $m_\text{H}$ the mass of the hydrogen atom. The density structure
is obtained from the equation of continuity. In these models we assume a
constant wind temperature $0.8T_\text{eff}$, and the electron density is
consistently calculated from the ionization balance. Since for these wind tests
we do not solve the equation of motion, it is necessary to specify wind
mass-loss rate for which the wind existence is tested. For each set of stellar
parameters the wind existence is tested for mass-loss rates
$10^{-8}\,\text{M}_\odot\,\text{year}^{-1}$,
$10^{-10}\,\text{M}_\odot\,\text{year}^{-1}$, and
$10^{-12}\,\text{M}_\odot\,\text{year}^{-1}$.

To obtain a correct surface flux we accounted for the Doppler effect during the
calculation of radiative transfer in lines, i.e. we shifted the stellar surface
flux according to the actual wind velocity \citep{babelb}.

%-----------------------------------------------------------------------
\subsection{Hydrodynamic wind models}

For those stars, for which the wind test showed that the wind is possible, we
also calculated hydrodynamic wind models \citepalias[described in][]{nltei}. In
these models we consistently solved the hydrodynamical equations, i.e. the
continuity equation, the momentum equation, and the energy equation with
radiative cooling and heating included using the electrons thermal balance
method \citep{kpp}.

These models enable us to derive the radial variations of wind density,
velocity, temperature, and the occupation numbers of individual excitation
states. They especially enable us to predict the wind mass-loss rate and the
terminal velocity.

%%%%%%%%%%%%%%%%%%%%%%%%%%%%%%%%%%%%%%%%%%%%%%%%%%%%%%%%%%%%%%%%%%%%%%%%
\section{Pure H-He winds}

In \citetalias{bezvi} we showed that the extremely massive ($M\geq100\,\ms$),
hot, pure hydrogen-helium stars do not have any wind. Stars very close to the
Eddington limit with $\Gamma\gtrsim0.859$ might be the only exception to this,
since these stars can have very weak pure hydrogen wind. The absence of winds is
connected with the material in the vicinity of these stars beeing ionized and
consequently the line transitions are not strong enough to drive a wind. On the
other hand, the material in the envelopes of the cooler, less massive stars
studied here may be less ionized, thus giving the possibility to drive a wind.
Here we test whether studied stars may have winds driven purely by hydrogen and
helium.

For very hot stars, with $\Teff \gtrsim5\cdot10^{4}\,\text{K}$ both hydrogen and
helium are completely ionized in the envelopes; consequently, the
radiative force due to the line transitions of these elements is by three to
four orders of magnitude lower than the absolute value of the gravitational
force. The most significant contribution to the radiative force is caused by the
light scattering on free electrons for these stars.

For cooler stars, helium may be partly singly ionized; however, these stars have
very broad photospheric lines, making the flux at the line positions so
low that it does not allow the wind to accelerate. For stars with
$T_\text{eff}\lesssim20\,000\,\text{K}$, helium may in addition become partly
neutral (in the envelopes with highest densities). However, the flux in the
ultraviolet part of the spectrum where the neutral helium resonance lines appear
is so low that the radiative force due to these lines is too low to accelerate
the wind. In all these envelopes the hydrogen is nearly completely ionized,
so that the radiative force due to hydrogen lines is negligible.

%%%%%%%%%%%%%%%%%%%%%%%%%%%%%%%%%%%%%%%%%%%%%%%%%%%%%%%%%%%%%%%%%%%%%%%%
\section{Analytical models of CNO line-driven winds}

Before discussing more detailed numerical NLTE wind models of CNO line-driven
winds, we first provide simplified analytic models of these winds
\citep[c.f.,][]{abb,owopo,feslo}.

The radiative acceleration in the Sobolev approximation
is the sum of the contributions by individual lines \citep{cassob}
\begin{equation}
\label{zarzrych}
g^{\text{rad}}=
  \frac{8\pi}{\rho c^2}\frac{v}{r} \sum_{\mathrm{lines}}
   \nu_{ij} H_c(\nu_{ij})\int_{\mu_*}^{1}\de\mu\,\mu\zav{1+\sigma\mu^2}
   \zav{1-e^{-\tau_\mu}},
\end{equation}
where $H_c(\nu_{ij})$ is the emergent flux from the stellar atmosphere at the
frequency $\nu_{ij}$ of a given line, $v$ is the radial wind velocity,
$\rho$ the wind density,
\begin{align}
\label{sigma}
\sigma=&\frac{r}{v}\frac{\de v}{\de r} -1,\\
\label{mu}
\mu_*=&\zav{1-\frac{R_*^2}{r^2}}^{1/2},
\end{align}
and the Sobolev optical depth $\tau_\mu$ is given by \citep{cassob,rybashumrem}
\begin{equation}
\label{tau}
\tau_\mu=\frac{\pi e^2}{m_\mathrm{e}\nu_{ij}}
\zav{\frac{n_i}{g_i}-\frac{n_j}{g_j}} g_if_{ij}
\frac{r}{v\zav{1+\sigma\mu^2}},
\end{equation}
where $n_i$, $n_j$, $g_i$, $g_j$ are the number densities and the statistical
weights of individual states giving rise to a given line with the oscillator
strength $f_{ij}$.

%=======================================================================
\subsection{Wind driven by optically thick lines}
\label{kapribod}
%stranka 6 - Praha neprijima a skoncili jsme v Drazdanech...

Generally, the stellar wind of present hot stars is driven by an ensemble of
optically thick and thin lines, the optically thick lines being important due to
their strength and the optically thin ones due to their large number
\citep[e.g.,][]{pusle}. The situation in the winds driven purely by CNO elements
is different. Contrary to say, iron, these elements do not have many
lines to drive wind, and the optically thick lines become much more important.
In many cases there are just few optically thick lines that drive a wind. In
such cases (for $\tau_\mu\gtrsim3$ to obtain a precision of $5\%$), the radiative
acceleration Eq.~\eqref{zarzrych} is given roughly by
\begin{equation}
\label{semtin}
g^{\text{rad}}_\text{thick}=
  \frac{4\pi v R_*^2}{\rho c^2 r^3} 
     \hzav{\frac{r}{v}\frac{\de v}{\de r}-
       \frac{1}{2}\zav{\frac{r}{v}\frac{\de v}{\de r} -1}\frac{R_*^2}{r^2}}
     \sum_{\tlustocerv}\nu_{ij} H_c(\nu_{ij}).
\end{equation}
The momentum equation 
neglecting the gas pressure term and introducing the Eddington parameter,
\begin{equation}
\label{sobcice}
\Gamma=\frac{g_\text{e}^\text{rad}}{g}=
\frac{1}{g}\frac{\sigma_\text{e} L}{4\pi c r^2},
\end{equation}
takes the form
\begin{equation}
\label{pardubice}
F(r,v,\frac{\de v}{\de r})=
v\frac{\de v}{\de r}-g^{\text{rad}}_\text{thick}+\frac{(1-\Gamma)GM}{r^2}=0.
\end{equation}
Here, $g_\text{e}^\text{rad}$ is the radiative acceleration due to the free
electrons, $L$ the stellar luminosity,
$\sigma_\text{e}={n_\text{e}s_\text{e}}/{\rho}$, $s_\text{e}$ the Thomson
scattering cross-section, $n_\mathrm{e}$ the electron number density, and
$g=GM/r^2$. The momentum equation \eqref{pardubice} has the critical point 
defined by
$\left.\partial F(r,v,\frac{\de v}{\de r})/\partial\zav{\frac{\de v}{\de r}}
\right|_{r_\text{crit}}=0$
($r_\text{crit}$ is the critical point radius)
\begin{equation}
\label{zamrsk}
\dot M\equiv 4\pi r^2 \rho v=
\frac{16\pi^2 R_*^2}{c^2}\zav{1-\frac{1}{2}\frac{R_*^2}{r_\text{crit}^2}}
 \sum_{\tlustocerv}\nu_{ij} H_c(\nu_{ij}),
\end{equation}
which yields the wind mass-loss rate. 
Contrary to the original analysis presented by \citet[hereafter CAK]{cak}, the
critical point condition is simpler, as the radiative force depends on the velocity
gradient linearly here.
It would be possible to use the regularity condition to derive $r_\text{crit}$
and $\mdot$ (as was done in CAK); however, as in most cases
$r_\text{crit}\approx R_*$, the wind mass-loss rate can be roughly given by
\begin{equation}
\label{brandejs}
\dot M \approx
\frac{8\pi^2 R_*^2}{c^2}\sum_{\tlustocerv}\nu_{ij} H_c(\nu_{ij}).
\end{equation}
This equation states that the mass-loss rate of the wind driven purely by
optically thick lines is approximately given by the photon mass-loss rate
($L/c^2$) multiplied by the number of thick lines 
\citep{lusol,zakladka}.
It can be shown that Eq.~\eqref{zamrsk} also defines the point where the wind
velocity is equal to the speed of so-called Abbott waves 
\citep{abb,fero}.

Using the mass-loss rate estimate Eq.~\eqref{brandejs} and the definition of the
mass-loss rate ($\dot M=4\pi r^2 \rho v$), the momentum equation
Eq.~\eqref{pardubice} can be rewritten as
\begin{equation}
\label{jedmomrov}
v\zav{1-\frac{R_*^2}{r^2}}\frac{\de v}{\de r}=
\frac{(1-\Gamma)GM}{r^2}-\frac{R_*^2 v^2}{r^3}.
\end{equation}
This equation has a critical point at $r_\text{crit}=R_*$.
To ensure the finiteness of the velocity derivative at
this point, the righthand side of Eq.~\eqref{jedmomrov}
should be equal to zero. This yields the value of the wind velocity at the
critical point
\begin{equation}
\label{kritrych}
v_\text{crit} =\sqrt{\frac{(1-\Gamma)GM}{R_*}}.
\end{equation}
The application of the l'H\^opital rule to the momentum equation
Eq.~\eqref{jedmomrov} at the critical point \citep{caslam} gives the value of
the velocity derivative at this point:
\begin{equation}
\label{rychde}
\frac{\de v}{\de r}=\frac{1}{4R_*}\sqrt{\frac{(1-\Gamma)GM}{R_*}}.
\end{equation}

%=======================================================================
\subsection{Theoretical limits for CNO line-driven winds}

In an accelerating outflow, the radiative force and the thermal pressure have to
overcome gravity and inertia at each point in wind. As the thermal pressure is
only effective at the base of the wind, the necessary condition for launching
line-driven stellar wind is that the magnitude of radiative force should exceed
the magnitude of gravitational force at a certain point in the wind,
Eq.~\eqref{trebova}. We apply this condition to derive the minimal mass-loss
rate of homogeneous wind (i.e. composed of all constituents, namely hydrogen,
helium, carbon, nitrogen, and oxygen).

For a given set of lines, the radiative force is maximum if the lines are not
self-shadowed, i.e.~if the lines are optically thin and the line optical depths
$\tau<1$ \citep{gayley}. Thus, for given occupation numbers, the line
acceleration Eq.~\eqref{zarzrych} cannot be higher than \citepalias[see
also][Eq.\,(2)]{bezvi}
\begin{equation}
\label{cerekvice}
g^\text{rad, max}=\frac{4\pi^2e^2}
{\rho m_\text{e}c^2}\zav{\frac{\rs}{r}}^2 \sum_{\text{lines}}
H_c (\nu_{ij}) g_if_{ij}\zav{\frac{n_i}{g_i}-\frac{n_j}{g_j}}.
\end{equation}
The condition of the wind existence Eq.~\eqref{trebova} then reads as
\cite[c.f.][]{kudmet}
\begin{equation}
\label{ostromer}
g^\text{rad, max}+g_\text{e}^\text{rad}=\zav{\bar Q+1}g_\text{e}^\text{rad}>g,
\end{equation}
where the
parameter $\bar Q\equiv g^\text{rad, max}/g_\text{e}^\text{rad}$ was introduced
by \citet{gayley}. Neglecting the term with the number density of the upper
level in Eq.~\eqref{cerekvice} (which means neglecting stimulated emission) and
substituting $g=GM/r^2$, the wind condition \eqref{ostromer} can be rewritten as
\begin{equation}
\label{kovac}
\frac{4\pi^2e^2R_*^2}{\rho m_\text{e}c^2 G{M}} 
 \sum_{\text{lines}} H_c (\nu_{ij}) f_{ij}n_i+\Gamma>1.
\end{equation}

%Ed tady nevim co chtela
%Kued: pridat pomlcku
For a very low density of CNO elements, only the strongest lines (i.e., the
resonance lines of the most abundant ions) significantly contribute to the
radiative force \citep [e.g.,][] {kkiv}. In such a case, condition
\eqref{kovac} can be rewritten as
\begin{equation}
\label{butoves}
\frac{4\pi^2e^2 R_*^2}{m_\text{e}c^2 G{M}}
\sum_{\text{res}} H_c (\nu_{ij}) f_{ij}\frac{Z_\text{el}}{m_\text{el}}+\Gamma>1,
\end{equation}
where $Z_\text{el}$ is the mass fraction of element corresponding to the
transition $i\leftrightarrow j$ (the ratio of the element density to the wind
density), $m_\text{el}$ is the element mass, and the summation goes over the
resonance lines of only the most abundant ions. If there is only one such
element, then the minimum mass fraction of this element needed to drive the
wind is given by
\begin{equation}
\label{jicin}
Z_\text{el}=\frac{m_\text{e}m_\text{el}c^2 G{M}}{4\pi^2e^2R_*^2}\zav{1-\Gamma}
\zav{\sum_{\text{res}} H_c (\nu_{ij}) f_{ij}}^{-1}.
\end{equation}
In scaled quantities, Eq.~\eqref{jicin} takes the form of
\begin{multline}
\label{vitineves}
Z_\text{el}=0.04
A_\text{el}\zav{1-\Gamma}\zav{\frac{M}{1\,\ms}}\zav{\frac{\rs}{1\,\text{R}_\odot}}^{-2}
\times \\*
\hzav{\sum_{\text{res}} \zav{\frac{H_c (\nu_{ij})}{10^{-7}\,\text{erg}\,
\text{cm}^{-2}}}f_{ij}}^{-1},
\end{multline}
where $A_\text{el}$ is the elemental mass in units of hydrogen mass.
Consequently, the minimum metallicity necessary for driving a wind depends on the
surface gravity $g=GM/R_*^2$, and on the stellar effective temperature that
determines both the ionization balance and the radiation flux.

The minimum mass-loss rate that can be driven from the star is given by the
condition that at least one line remains to be optically thick at the critical
point, i.e. $\tau_\mu>1$, where $\tau_\mu$ is given by Eq.~\eqref{tau}. Using
Eqs.~\eqref{kritrych} and \eqref{rychde}, we can show (assuming $r_\text{crit}
\approx R_*$ and for radial rays $\mu=1$) that the mass-loss rate should fulfil
the condition
\begin{equation}
\label{minztrata}
\dot M>\frac{m_\text{el}m_\text{e}}{e^2}\frac{\nu_{ij}(1-\Gamma)GM}
{f_{ij}Z_\text{el}}.
\end{equation}
For a typical case where there is only one optically thick line left, it can be
readily shown (using the mass-loss rate estimate Eq.~\eqref{brandejs}) that the
condition Eq.~\eqref{minztrata}
\begin{equation}
\label{butovesstej}
\frac{ 8 \pi^2 e^2 R_*^2Z_\text{el}}{m_\text{e}m_\text{el}c^2 G{M}}
H_c (\nu_{ij}) f_{ij}+\Gamma>1,
\end{equation}
is nearly the same as the condition Eq.~\eqref{butoves} written for one line.
Consequently, if there is a line capable of overcoming the gravity (according to
Eq.~\eqref{butoves}), it is automatically optically thick one, and this
line sets the mass-loss rate (and vice versa).
In other words, for very low-density winds the condition of maximum radiative
force Eq.~\eqref{ostromer} and the critical point analysis Sect.~\ref{kapribod}
give the same results
for the existence of the wind
and the wind's mass-loss rate.

%%%%%%%%%%%%%%%%%%%%%%%%%%%%%%%%%%%%%%%%%%%%%%%%%%%%%%%%%%%%%%%%%%%%%%%%
\section{CNO-driven wind models}

\begin{figure}
\centering
\resizebox{0.9\hsize}{!}{\includegraphics{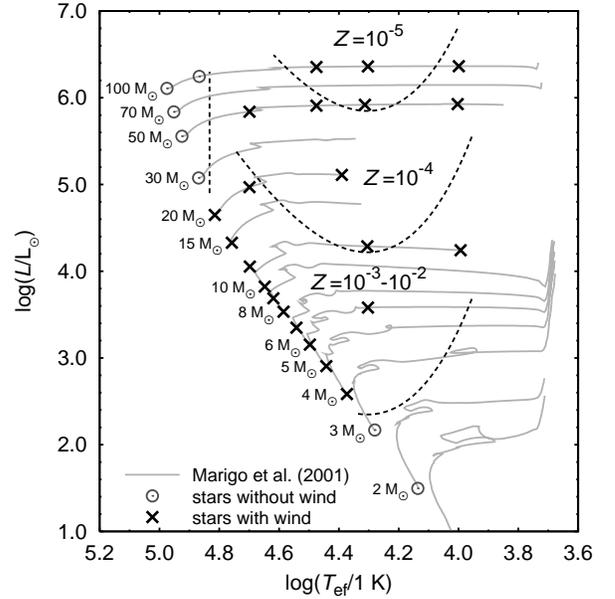}}
\caption{The HR diagram with minimal metallicity needed to drive a wind.
Stars for which the winds are possible (for a studied range of metallicities)
are denoted by crosses, and stars for which the winds are not possible are
denoted by circles. The dashed lines denote boundaries separating
stars with a different minimum metallicity necessary for driving a wind. The
boundaries were derived using Eq.~\eqref{butoves} and the Wien law. Overplotted
are the evolutionary tracks of \citet{bezmari}.}
\label{hrdrov}
\end{figure}

For the parameters of model stars given in Table~\ref{hvezpar}, we tested whether
these stars could have CNO driven winds. We tested the possibility of the wind
for the mass-loss rates $10^{-12}\,\msr$, $10^{-10}\,\msr$, and $10^{-8}\,\msr$,
and for the mass fraction of CNO elements in the range of $10^{-4}-0.1$ with a
relative step of 10. We assumed the same number density for carbon, nitrogen,
and oxygen.

The regions in the HR diagram corresponding to the parameters of stars for which
the wind tests are positive are given in Fig.~\ref{hrdrov}. The minimum mass
fraction of CNO needed to drive the wind is also indicated there.

For parameters for which the wind may exist, we calculated 
hydrodynamic wind models and
predicted the mass-loss rate (again assuming the same number density of CNO).
Calculated wind parameters for individual stars are given in Appendix in
Table~\ref{spoctene}. The mass-loss rate of studied stars can be fitted by the
formula
\begin{equation}
\label{fitrov}
\dot M(L,Z,T_\text{eff})= \alpha_0\, L^{\alpha_1}\,
10^{\textstyle\frac{\alpha_2\zav{\log Z +\alpha_3\log L}}
                   {\log Z(\alpha_4+\alpha_5\log L)+1}
    \scriptstyle+
\alpha_6\log T_\text{eff}},
\end{equation}
where the luminosity $L$ is expressed in the solar luminosity units
$\text{L}_\odot$, $Z$ is the mass fraction of heavier elements, and
$T_\text{eff}$ is expressed in the units of Kelvin. The values of parameters
$\alpha_0, \dots, \alpha_6$ fitted using the subroutine VARPRO
\citep{varpro1,varpro2} for the stellar groups with different effective
temperatures are given in Table~\ref{fittab}. For $\log Z(\alpha_4+\alpha_5\log
L)+1<0$ the predicted mass-loss rate is set to zero.

\begin{figure}
\centering
\resizebox{0.8\hsize}{!}{\includegraphics{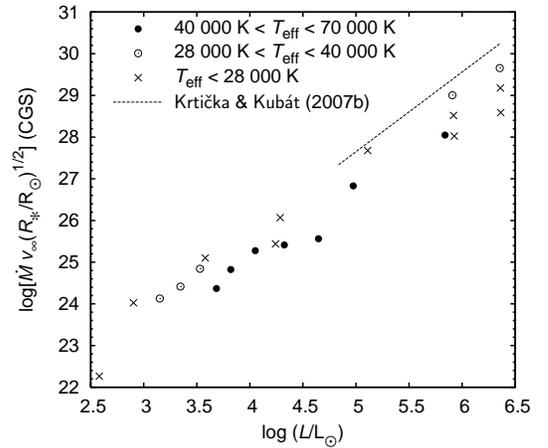}}
\caption{The modified wind momentum-luminosity relationship for stars with
different effective temperatures and $Z=0.01$. Stars with
$T_\text{eff}\gtrsim70\,000\,\text{K}$ do not have any wind.
Overplotted is the mean dependence of the modified wind momentum predicted for O
stars with solar metallicity by \citet{spravnez}.}
\label{mom_porov}
\end{figure}

In present-day hot stars, the modified wind momentum $\mdot \vinfty
\zav{R/\text{R}_{\sun}}^{1/2}$ depends mainly on the stellar luminosity
\citep[and references therein]{kupul}. According to its value, these stars
can be divided into four groups (see Fig.~\ref{mom_porov}).

\begin{table*}[hbt]
\caption{The best-fit parameters of Eq.~\eqref{fitrov} for individual stellar
groups}
\label{fittab}
\centering
\begin{tabular}{cccccccc}
\hline
$T_\text{eff}$ [K]  & $\alpha_0$ [$\msr$]
& $\alpha_1$ & $\alpha_2$ & $\alpha_3$ & $\alpha_4$
& $\alpha_5$ & $\alpha_6$ \\
\hline
$<28\,000$ & $1.19\times10^{-23}$ & $0.8437$ & $0.2936$ & $0.7167$ & $0.1208$ &
$0.0064$ & $2.4369$  \\
$28\,000-40\,000$ & $8.92\times10^{-11}$ & $0.7895$ & $0.6123$ & $0.8181$ & 
$0.2848$ & $-0.020158$ & $-0.7394$ \\
$>40\,000$ & $2.39\times10^{11}$ & $0.5722$ & $0.8348$ & $0.7995$ & $0.3983$ &
$-0.03542$ & $-5.363$\\
\hline
\end{tabular}
\end{table*}

%=======================================================================
\subsection{Stars with no wind: $T_\text{eff}\gtrsim70\,000\,\text{K}$}

For the hottest stars with the effective temperature
$T_\text{eff}\gtrsim70\,000\,\text{K}$, we did not find that the wind exists.
Although these stars are very luminous, their envelope is highly ionized with
ions \ion{}{vi}--\ion{}{vii} being the dominant ionization states of CNO
elements. Since the factor
$\zav{{M}/{1\,\ms}}\zav{{\rs}/{1\,\text{R}_\odot}}^{-2}$ in
%Ed myslis napsat cislovku?
Eq.~\eqref{vitineves} is
%Kued: higher
larger
than one for all these stars, a strong flux at
the line position is needed to accelerate the wind. However, ions
\ion{}{vi}--\ion{}{vii} have strong resonance lines in the far UV region (for
frequencies higher than the \ion{He}{ii} ionization threshold), where very low flux
is emitted, so we do not find any wind for these stars.

%=======================================================================
\subsection{Hottest stars with winds: $40\,000\,\text{K}\lesssim
T_\text{eff}\lesssim70\,000\,\text{K}$}

Stars with the effective temperatures in the range $40\,000\,\text{K}\lesssim
T_\text{eff}\lesssim70\,000\,\text{K}$ may have a line-driven wind accelerated
mainly by the \ion{O}{v} and \ion{O}{vi} lines. The hydrogen-like ion
\ion{C}{vi} and the helium-like one \ion{N}{vi}, which are dominant ionization
fractions of carbon and nitrogen, do not contribute significantly to the
radiative force. The latter ions do not have enough strong resonance lines at
the wavelengths at which strong flux is emitted. The minimum mass fraction of
%Ed trochu upraveno
CNO elements needed to drive a wind agrees with analytical estimate
Eq.~\eqref{vitineves} and is relatively high, of the order of $10^{-3}$ for ZAMS
stars and as low as $10^{-4}$ for more evolved stars with larger radii.

The winds of hottest stars in this sample are accelerated mainly by the
\ion{O}{vi} resonance doublet 2s\,\element[\text{e}][2]{S} $\rightarrow$
2p\,\element[\text{o}][2]{P}. The analytic formula Eq.~\eqref{brandejs} gives
the asymptotic value of the mass-loss rate if both lines of the doublet are
optically thick (i.e., for high metallicity), while for lower metallicity the
lines become optically thinner and the mass-loss rate decreases. For example,
for the model star M150, the agreement between the mass-loss rate given by
Eq.~\eqref{brandejs} $ 7.5 \times10^{-10}\,\msr$ and the mass-loss rate
$6.4\times10^{-10}\,\msr$ predicted for $Z=0.02$ by our hydrodynamic wind models
shows a reliability of Eq.~\eqref{brandejs}.

The winds of ZAMS stars are very weak (with the mass-loss rate of the order of
$10^{-10}\,\msr$ for the solar chemical composition) due to their relatively low
luminosity, whereas the winds of supergiants may be stronger with the mass-loss
close to $10^{-7}\,\msr$.

For these stars the fitting formula Eq.~\eqref{fitrov} only gives reliable
predictions for $\dot M\gtrsim10^{-9}\,\msr$, for winds with smaller $\dot
M$ the error could be of a factor of 3.

%=======================================================================
\subsection{Stars with $28\,000\,\text{K}\lesssim
T_\text{eff}\lesssim40\,000\,\text{K}$}

For stars $28\,000\,\text{K}\lesssim T_\text{eff}\lesssim40\,000\,\text{K}$, the
wind is possible, and carbon and nitrogen become more important for the driving
of the wind, especially for cooler stars. The metallicity needed to drive a
wind is about $10^{-4}-10^{-3}$ for ZAMS stars and may be extremely low
($\sim10^{-5}$) for most massive evolved stars. This is connected with
the relatively large factor
$\zav{{M}/{1\,\ms}}\zav{{\rs}/{1\,\text{R}_\odot}}^{-2}$ in
Eq.~\eqref{vitineves} for ZAMS stars (these stars are
relatively compact), whereas this factor is lower for massive evolved stars with
large radii.

%=======================================================================
\subsection{The coolest stars with $10\,000\,\text{K}\lesssim
T_\text{eff}\lesssim28\,000\,\text{K}$}

The coolest ZAMS stars with $T_\text{eff}\lesssim20\,000\,\text{K}$ do not have
%Ed to je nejake divne
any wind because these stars have
%Kued:
a
too high surface gravity to accelerate the
wind. On the other hand, evolved massive stars may have winds even for metallicities as low
as $10^{-5}$. For $Z\approx0.01$, Pop III supergiant winds may have
relatively high mass-loss rates of about $10^{-6}\,\msr$, comparable to
present-day hot star winds. The winds of these stars are mainly accelerated by
singly and doubly ionized carbon and nitrogen. Oxygen does not 
contribute significantly to the radiative force in this temperature range as the resonance
lines of singly and doubly ionized oxygen have higher frequencies than the
hydrogen ionization frequency, and only a relatively weak radiative flux is
emitted there.

For the coolest stars, the fitting formula Eq.~\eqref{fitrov} can only be used
for higher mass-loss rates $\dot M\gtrsim10^{-9}\,\msr$.
%%%%%%%%%%%%%%%%%%%%%%%%%%%%%%%%%%%%%%%%%%%%%%%%%%%%%%%%%%%%%%%%%%%%%%%%
\section{Consequences of the wind's existence}

\subsection{Influence on the stellar evolution}

Our results show that the winds of first stars driven by CNO elements are rather
weak compared to the winds of present-day hot stars. This is caused by two
effects, the smaller number of CNO lines compared to, e.g., iron ones, and
by the large surface gravity of first stars. Consequently, the estimated total mass
lost due to winds during the stellar evolution is rather small. For example,
assuming that the final abundance of CNO elements is a solar one, the estimate
of the total mass loss of a star with an initial mass $M_0=100\,\ms$ is arround
$1\,\ms$, and that of a star with $M_0=20\,\ms$ is about $0.1\,\ms$.
Consequently, such stars lose only up to a few percent of their mass via
line-driven winds.

In reality, the total mass lost might be even lower because CNO elements
may occur at the stellar surface only in the later phases of the stellar
evolution and the surface metallicity might be lower than the solar one.

%=======================================================================
\subsection{Metal enrichment of the primordial halos}

The presence of a relatively small amount of metals (with
$Z\approx10^{-5}-10^{-4}$) in the primordial haloes might inhibit subsequent
formation of supermassive stars \citep[see][for a review]{lesdiablerets}.
Consequently, if the stellar winds of massive stars are strong enough, they can
change the initial mass function in the primordial haloes.

If in the halo with the baryonic mass of about $10^5\,\ms$ a first star forms
with the mass of $100\,\ms$, it might lose mass arround $1\,\ms$ due to
the line-driven winds. Consequently, roughly $10^{-2}\,\ms$ of
freshly synthesized CNO elements occur in such a halo, leading to the mass fraction of heavier
elements of about $10^{-7}$, which is too low to change the initial mass
function. The metallicity of the primordial halo would be enhanced more
significantly only after the considered star explodes as a supernova.

%%%%%%%%%%%%%%%%%%%%%%%%%%%%%%%%%%%%%%%%%%%%%%%%%%%%%%%%%%%%%%%%%%%%%%%%
\section{Discussion}

\subsection{The chemical composition}

The mass-loss rate predictions were calculated for the same number density of
each of the CNO elements. In reality, the fraction of these elements is not
uniform as assumed here. In such a case an approximate estimate of the wind
mass-loss rate can be obtained as for different temperatures the wind is
accelerated by different elements. For the hottest stars with
$T_\text{eff}\gtrsim 40\,000\,\text{K}$, the wind is accelerated mainly due to
oxygen lines (as supported by our models driven purely by oxygen), consequently
for these stars the mass-loss rate depends mainly on the oxygen abundance. For
%Ed pridal jsem of
%Kued: chtela to
cooler stars the wind is accelerated because of both carbon and nitrogen,
so it could be roughly assumed that the wind mass-loss rate scales
either with carbon or nitrogen abundance.

%=======================================================================

Stellar winds of present {\em luminuous} hot stars are mainly accelerated thanks to
numerous heavier element lines \citep[especially of iron and also of,
e.g.,~nickel or copper, e.g.,][]{pasam,pusle,vikolamet} in the region where the
wind mass-loss rate is determined, i.e. below the wind critical point. CNO
elements do not have many strong lines available to drive a wind
%Ed uprava
as do heavier elements. In the formalism of line-strength distribution functions, CNO
elements and iron-peak elements dominate in different parts of the line-strength
distribution function \citep{pusle}. As the result of this,
CNO elements in present hot stars are important in the case when their large abundance (compared to
iron-peak elements) enables their lines to continue beeing optically thick, i.e.,
in the outer wind regions of luminuous stars or in the weak winds of less
luminuous stars \citep[e.g.,][]{vikolamet,nlteii,kkiv}. These trends are
reflected in
the CNO mass-loss rate predictions presented here. Presented mass-loss rate
predictions of luminuous stars are significantly lower than those derived
assuming solar mixture of heavier elements
\citep[e.g.,][]{vikolamet,kudmet,nltei,graham}. On the other hand, the pure CNO
mass-loss rate of less luminuous stars (or stars with very low metallicity)
corresponds roughly to what is calculated assuming solar mixture of elements.

%=======================================================================
\subsection{Comparison with literature}

Since the wind models driven purely by CNO elements were not, to our knowledge,
calculated for the studied stars, a direct comparison between our results and
the results available in the literature is not possible.
%Ed trochu jinak
Pure CNO wind models
were calculated by \citet{uncno}, but for different stellar parameters. However, at a
very low wind density, the stellar winds are accelerated mostly by CNO lines and
the contribution of heavier elements is basically negligible
\citep[e.g.,][]{vikolamet,kkiv}. Consequently, an indirect comparison between
our predictions and the predictions available in the literature for the lowest
metallicities is feasible.

\citet{kudmet} provided mass-loss rate predictions for very massive O stars at
very low metallicities. The selected stellar sample overlaps with ours for the
stars with mass $100\,\ms$. For these stars the lowest metallicities for which
\citet{kudmet} provides the mass-loss rate predictions reasonably agree with our
minimum values of the metallicity needed to drive a wind (see Fig.~\ref{hrdrov}).
There is also reasonable agreement between our models and the models with the
lowest mass-loss rate calculated by \citet{vikolamet} and \citet{vikowr}.

%=======================================================================
\subsection{Multicomponent effects and pure metallic wind}

The hydrogen and helium components of the stellar wind are accelerated mainly
because of the collisions of hydrogen and helium atoms with heavier elements.
Consequently, stellar winds of hot stars have a multicomponent nature
\citep[e.g.,][]{op,nlteii}. For low-density winds the transfer of momentum
between radiationally accelerated metals and hydrogen and helium becomes
inefficient. In this case the frictional heating or the decoupling of wind
components may influence the wind structure.

If the decoupling occurs for velocities that exceed the escape speed, then the
multicomponent effects do not influence the mass-loss rate \citep[e.g.,][]{ufo}.
On the other hand, if the decoupling occurs close to the star or in the stellar
atmosphere, then the existence of pulsating shells \citep{obalka} or pure
metallic wind \citep{babelb} is possible.

Following \citet{nlteii}, we calculated the nondimensional velocity difference
$x_{h\text{p}}$ between a given heavier element $h$ and the hydrogen-helium
component. For small velocity differences $x_{h\text{p}}\lesssim0.1$, the wind is
well-coupled and can be modelled as a one-component one. For larger velocity
differences $0.1\lesssim x_{h\text{p}}\lesssim1$, the frictional heating
influences the wind temperature, and for large velocity differences
$x_{h\text{p}}\gtrsim1$, the wind components decouple. The tests showed that, for
nearly all wind models of ZAMS stars and for the low metallicity models of
evolved stars, the multicomponent effects are indeed important.  However, in most
cases the multicomponent effects start to be important in the outer wind
regions, so these effects do not affect the wind mass-loss rate.

For the cases when homogeneous wind is not possible, a pure metallic wind may
exist \citep{babelb}. We postpone the study of such stellar winds to a separate
paper.

%=======================================================================
\subsection{Model simplifications}

The radiative force in the models presented here is calculated neglecting line
overlaps and the influence of the line transitions on the continuum radiative
transfer \citep[see][for a more detailed discussion of the simplification of our
code]{nltei}. However, these effects do not significantly influence the results
here. First, CNO elements do not have as many lines as
%Ed a je divne
%Kued: to je, zkusim to jinak
iron-peak elements; consequently, the line overlaps
%Kued: do not play as significant role in CNO winds
in CNO winds do not play as significant role
as in the winds driven also by heavier elements. Second, the
calculated mass-loss rate is relatively low, so the number of strong
optically thick lines is relatively even lower. Finally, even for the winds
driven by the solar mixture of elements our models are able to predict reliable
wind parameters.

Hot star winds display small-scale inhomogeneities (clumping) that may influence
the predicted wind mass-loss rate \citep[e.g.,][]{lijana}. As the effect of
these inhomogeneities on the wind mass-loss rate is not yet known, we neglect
this effect.

%=======================================================================
\subsection{H-He winds and other ways how to lose mass}

A question might appear as to why the stellar wind driven purely by hydrogen and
helium lines is not possible. This question is even more appealing in view
of the fact that, in some models, the radiative force due to hydrogen may
contribute a few percent to the total radiative force. To obtain a wind, in
such a case, one would have to decrease the wind density. However, as the neutral
hydrogen originates from the recombination, the decrease in the wind density
leads to the decrease in the neutral hydrogen fraction and, consequently, to the
decrease in the radiative force due to hydrogen. Consequently, as a result of
this coupling the winds driven purely by hydrogen are not possible. A similar
effects also occur for helium.

%=======================================================================

We note that there are several other possibilities how massive stars can lose
their mass, although these possibilities were not up to now studied in greater
detail. Stars with the luminosity higher than the corresponding Eddington limit
may lose mass due to porous winds \citep{owosha} or via $\eta$~Car type of
explosions \citep{vodovar}. Moreover, stars rotating with the critical rotation
rate may lose their mass via equatorial disc \citep[e.g.,][]{mee}.

%%%%%%%%%%%%%%%%%%%%%%%%%%%%%%%%%%%%%%%%%%%%%%%%%%%%%%%%%%%%%%%%%%%%%%%%
\section{Conclusions}

We have studied the stellar winds of massive first stars to show that pure
hydrogen-helium, hot first stars do not have any stellar wind. Only stars very
close to the Eddington limit may have very weak pure hydrogen wind
\citepalias{bezvi}. 

As soon as CNO elements are synthesized in the stellar core and transported to
the surface via the mixing processes, 
the CNO driven wind may exist for stars that fulfil the wind condition Eq.~\eqref{butoves}. We calculated the models for
this wind and provided an approximate formula for calculating the mass-loss rate as
a function of stellar parameters.

With decreasing $Z$, both the mass-loss rate and the number of optically thick
lines decrease. We have shown that the condition that at least one line is
optically thick Eq.~\eqref{butovesstej} is nearly the same as the wind
condition, Eq.~\eqref{butoves} according to which the radiative force should be
greater than gravity. Consequently, the wind ceases to
exist for decreasing $Z$ both because the radiative force is too low and
because all lines become optically thin.

We discussed the influence of CNO driven stellar winds on the stellar evolution
and circumstellar environment. As the CNO elements are not able to
accelerate the wind as efficiently as heavier elements, the wind mass-loss rates
of CNO winds are relatively low. Consequently, the CNO winds do not
significantly influence the stellar evolution or the circumstellar and
interstellar environments.

We conclude that line-driven winds strong enough to influence the
evolution of stars far away from the Eddington limit did not occur before the
first supernova explosions.

\begin{acknowledgements}
We thank Dr. A. Feldmeier for his comments on the manuscript.
This research made use of NASA's ADS, and the NIST database {\sf
http://physics.nist.gov/asd3}. This work was supported by grant GA \v{C}R
205/07/0031. The Astronomical Institute Ond\v{r}ejov is supported by the project
AV0\,Z10030501.
\end{acknowledgements}

\newcommand{\actob}{Active OB-Stars:
	Laboratories for Stellar \& Circumstellar Physics, 
 	eds. S. \v{S}tefl, S. P. Owocki, \& A.~T.
        Okazaki (San Francisco: ASP Conf. Ser)}

\newcommand{\pulrot}{Pulsation, rotation, and mass loss in early-type stars,
IAUS 162, eds. L. A. Balona, H. F. Henrichs, \& J. M. Contel. (Dordrecht: Kluwer
Academic Publishers)}

\appendix

\section{The derived wind parameters}
\onecolumn

\begin{table}[tbh]
\caption{The wind parameters of studied stars calculated using
hydrodynamic models}
\label{spoctene}
\begin{center}
\begin{tabular}{lccc}
\hline
Model & $Z$ & $\mdot$ & $v_\infty$ \\
 & & [$\text{M}_\odot\,\text{year}^{-1}$] & [$\text{km}\,\text{s}^{-1}$]\\
\hline
\input{pisvse.tex}
\hline
\end{tabular}\hspace{1cm}
\begin{tabular}{lccc}
\hline
Model & $Z$ & $\mdot$ & $v_\infty$ \\
 & & [$\text{M}_\odot\,\text{year}^{-1}$] & [$\text{km}\,\text{s}^{-1}$]\\
\hline
\input{pisvsf.tex}
\hline
\end{tabular}
\end{center}
\end{table}

\end{document}

%% file: pisvse.tex
M040 & $1\times10^{-2}$ & $2.9\times10^{-12}$ & $950$ \\
M040 & $2\times10^{-2}$ & $1.2\times10^{-11}$ & $1520$ \\
M050 & $3\times10^{-3}$ & $1.4\times10^{-11}$ & $1010$ \\
M050 & $5\times10^{-3}$ & $3.5\times10^{-11}$ & $1430$ \\
M050 & $1\times10^{-2}$ & $7.1\times10^{-11}$ & $2120$ \\
M050 & $2\times10^{-2}$ & $6.7\times10^{-11}$ & $3220$ \\
M050-1 & $1\times10^{-3}$ & $1.3\times10^{-10}$ & $1100$ \\
M050-1 & $3\times10^{-3}$ & $2.9\times10^{-10}$ & $1570$ \\
M050-1 & $1\times10^{-2}$ & $7.2\times10^{-10}$ & $1220$ \\
M050-1 & $2\times10^{-2}$ & $1.2\times10^{-09}$ & $830$ \\
M060 & $3\times10^{-3}$ & $1.8\times10^{-11}$ & $1630$ \\
M060 & $5\times10^{-3}$ & $3.5\times10^{-11}$ & $2260$ \\
M060 & $1\times10^{-2}$ & $5.8\times10^{-11}$ & $3240$ \\
M060 & $2\times10^{-2}$ & $7.9\times10^{-11}$ & $4470$ \\
M070 & $5\times10^{-3}$ & $3.4\times10^{-11}$ & $2360$ \\
M070 & $1\times10^{-2}$ & $1.1\times10^{-10}$ & $3410$ \\
M070 & $2\times10^{-2}$ & $2.0\times10^{-10}$ & $4310$ \\
M080 & $3\times10^{-3}$ & $6.7\times10^{-11}$ & $2440$ \\
M080 & $5\times10^{-3}$ & $1.3\times10^{-10}$ & $3250$ \\
M080 & $1\times10^{-2}$ & $2.1\times10^{-10}$ & $4540$ \\
M080 & $2\times10^{-2}$ & $3.3\times10^{-10}$ & $5950$ \\
M090 & $5\times10^{-3}$ & $3.0\times10^{-11}$ & $3010$ \\
M090 & $1\times10^{-2}$ & $9.1\times10^{-11}$ & $3500$ \\
M090 & $2\times10^{-2}$ & $3.5\times10^{-10}$ & $4790$ \\
M100 & $3\times10^{-3}$ & $3.7\times10^{-11}$ & $2540$ \\
M100 & $5\times10^{-3}$ & $7.3\times10^{-11}$ & $3320$ \\
M100 & $1\times10^{-2}$ & $2.7\times10^{-10}$ & $3340$ \\
M100 & $2\times10^{-2}$ & $6.0\times10^{-10}$ & $3790$ \\
M100-1 & $1\times10^{-4}$ & $9.1\times10^{-11}$ & $420$ \\
M100-1 & $3\times10^{-4}$ & $5.4\times10^{-10}$ & $930$ \\
M100-1 & $1\times10^{-3}$ & $1.3\times10^{-09}$ & $1530$ \\
M100-1 & $3\times10^{-3}$ & $3.3\times10^{-09}$ & $1030$ \\
M100-1 & $1\times10^{-2}$ & $6.3\times10^{-09}$ & $870$ \\
M100-1 & $2\times10^{-2}$ & $7.5\times10^{-09}$ & $1330$ \\
M100-2 & $1\times10^{-2}$ & $8.1\times10^{-10}$ & $800$ \\
M100-2 & $2\times10^{-2}$ & $1.0\times10^{-09}$ & $1070$ \\
M120 & $1\times10^{-3}$ & $6.9\times10^{-12}$ & $1020$ \\
M120 & $2\times10^{-3}$ & $1.2\times10^{-10}$ & $2050$ \\
M120 & $3\times10^{-3}$ & $2.1\times10^{-10}$ & $2540$ \\
M120 & $5\times10^{-3}$ & $3.5\times10^{-10}$ & $3630$ \\
M120 & $1\times10^{-2}$ & $4.9\times10^{-10}$ & $5130$ \\
M120 & $2\times10^{-2}$ & $6.1\times10^{-10}$ & $7110$ \\
M150 & $1\times10^{-3}$ & $1.6\times10^{-11}$ & $1420$ \\
M150 & $2\times10^{-3}$ & $1.6\times10^{-10}$ & $2700$ \\
M150 & $3\times10^{-3}$ & $2.8\times10^{-10}$ & $3370$ \\
M150 & $5\times10^{-3}$ & $4.2\times10^{-10}$ & $4350$ \\
M150 & $1\times10^{-2}$ & $5.7\times10^{-10}$ & $5950$ \\
M150 & $2\times10^{-2}$ & $6.4\times10^{-10}$ & $8110$ \\
M200 & $3\times10^{-3}$ & $3.8\times10^{-10}$ & $3120$ \\
M200 & $5\times10^{-3}$ & $5.7\times10^{-10}$ & $4200$ \\
M200 & $1\times10^{-2}$ & $7.7\times10^{-10}$ & $5840$ \\
M200 & $2\times10^{-2}$ & $9.1\times10^{-10}$ & $7680$ \\
M200-1 & $3\times10^{-4}$ & $6.7\times10^{-10}$ & $1470$ \\
M200-1 & $1\times10^{-3}$ & $3.0\times10^{-09}$ & $2800$ \\
M200-1 & $3\times10^{-3}$ & $5.6\times10^{-09}$ & $4020$ \\
M200-1 & $1\times10^{-2}$ & $9.4\times10^{-09}$ & $5630$ \\
M200-1 & $2\times10^{-2}$ & $1.3\times10^{-08}$ & $5130$ \\

%% file: pisvsf.tex
M200-2 & $1\times10^{-4}$ & $7.0\times10^{-09}$ & $1000$ \\
M200-2 & $3\times10^{-4}$ & $2.0\times10^{-08}$ & $660$ \\
M200-2 & $1\times10^{-3}$ & $4.2\times10^{-08}$ & $780$ \\
M200-2 & $3\times10^{-3}$ & $5.5\times10^{-08}$ & $1490$ \\
M200-2 & $1\times10^{-2}$ & $6.4\times10^{-08}$ & $2650$ \\
M200-2 & $2\times10^{-2}$ & $6.9\times10^{-08}$ & $3360$ \\
M500-1 & $1\times10^{-4}$ & $8.3\times10^{-09}$ & $1510$ \\
M500-1 & $3\times10^{-4}$ & $2.7\times10^{-08}$ & $2440$ \\
M500-1 & $1\times10^{-3}$ & $6.3\times10^{-08}$ & $2750$ \\
M500-1 & $3\times10^{-3}$ & $9.9\times10^{-08}$ & $3460$ \\
M500-1 & $1\times10^{-2}$ & $1.3\times10^{-07}$ & $4060$ \\
M500-1 & $2\times10^{-2}$ & $1.6\times10^{-07}$ & $4680$ \\
M500-2 & $1\times10^{-4}$ & $7.6\times10^{-08}$ & $1010$ \\
M500-2 & $3\times10^{-4}$ & $1.4\times10^{-07}$ & $1450$ \\
M500-2 & $1\times10^{-3}$ & $4.0\times10^{-07}$ & $1930$ \\
M500-2 & $3\times10^{-3}$ & $5.9\times10^{-07}$ & $2640$ \\
M500-2 & $1\times10^{-2}$ & $8.1\times10^{-07}$ & $3400$ \\
M500-2 & $2\times10^{-2}$ & $9.5\times10^{-07}$ & $3580$ \\
M500-3 & $1\times10^{-5}$ & $4.7\times10^{-09}$ & $300$ \\
M500-3 & $3\times10^{-5}$ & $2.1\times10^{-08}$ & $650$ \\
M500-3 & $1\times10^{-4}$ & $4.7\times10^{-08}$ & $1040$ \\
M500-3 & $3\times10^{-4}$ & $9.7\times10^{-08}$ & $990$ \\
M500-3 & $1\times10^{-3}$ & $2.1\times10^{-07}$ & $580$ \\
M500-3 & $3\times10^{-3}$ & $2.8\times10^{-07}$ & $960$ \\
M500-3 & $1\times10^{-2}$ & $3.4\times10^{-07}$ & $1790$ \\
M500-3 & $2\times10^{-2}$ & $3.7\times10^{-07}$ & $2480$ \\
M500-4 & $1\times10^{-4}$ & $1.1\times10^{-09}$ & $160$ \\
M500-4 & $3\times10^{-4}$ & $2.0\times10^{-08}$ & $330$ \\
M500-4 & $1\times10^{-3}$ & $3.8\times10^{-08}$ & $600$ \\
M500-4 & $3\times10^{-3}$ & $5.0\times10^{-08}$ & $970$ \\
M500-4 & $1\times10^{-2}$ & $6.3\times10^{-08}$ & $1540$ \\
M500-4 & $2\times10^{-2}$ & $7.2\times10^{-08}$ & $1900$ \\
M999-1 & $1\times10^{-5}$ & $1.5\times10^{-08}$ & $350$ \\
M999-1 & $3\times10^{-5}$ & $1.6\times10^{-07}$ & $790$ \\
M999-1 & $1\times10^{-4}$ & $4.0\times10^{-07}$ & $1190$ \\
M999-1 & $3\times10^{-4}$ & $6.1\times10^{-07}$ & $1460$ \\
M999-1 & $1\times10^{-3}$ & $1.7\times10^{-06}$ & $2010$ \\
M999-1 & $3\times10^{-3}$ & $2.3\times10^{-06}$ & $2620$ \\
M999-1 & $1\times10^{-2}$ & $3.3\times10^{-06}$ & $2930$ \\
M999-1 & $2\times10^{-2}$ & $3.9\times10^{-06}$ & $2870$ \\
M999-2 & $1\times10^{-5}$ & $3.9\times10^{-08}$ & $430$ \\
M999-2 & $3\times10^{-5}$ & $1.1\times10^{-07}$ & $710$ \\
M999-2 & $1\times10^{-4}$ & $2.3\times10^{-07}$ & $860$ \\
M999-2 & $3\times10^{-4}$ & $4.6\times10^{-07}$ & $720$ \\
M999-2 & $1\times10^{-3}$ & $7.5\times10^{-07}$ & $650$ \\
M999-2 & $3\times10^{-3}$ & $8.9\times10^{-07}$ & $1210$ \\
M999-2 & $1\times10^{-2}$ & $1.0\times10^{-06}$ & $2050$ \\
M999-2 & $2\times10^{-2}$ & $1.3\times10^{-06}$ & $2130$ \\
M999-3 & $1\times10^{-4}$ & $3.7\times10^{-08}$ & $310$ \\
M999-3 & $3\times10^{-4}$ & $8.6\times10^{-08}$ & $450$ \\
M999-3 & $1\times10^{-3}$ & $1.3\times10^{-07}$ & $710$ \\
M999-3 & $3\times10^{-3}$ & $1.6\times10^{-07}$ & $1020$ \\
M999-3 & $1\times10^{-2}$ & $1.9\times10^{-07}$ & $1450$ \\
M999-3 & $2\times10^{-2}$ & $2.3\times10^{-07}$ & $1490$ \\